# Z-SEP: Zonal-Stable Election Protocol for Wireless Sensor Networks

S. Faisal[1], N. Javaid[1], A. Javaid[2], M. A. Khan[1], S. H. Bouk[1], Z. A. Khan[3]

[1]COMSATS Institute of Information Technology, Islamabad, Pakistan.
[2] Mathematics Dept., COMSATS Institute of Information Technology, Wah Cant., Pakistan.
[3]Faculty of Engineering, Dalhousie University, Halifax, Canada.
Email: nadeemjavaid@comsats.edu.pk, Phone: +92519049323

___________________________________________________________________________

**ABSTRACT**

Wireless Sensor Networks (WSNs) are comprised of thousands of sensor nodes, with restricted energy, that co-operate to accomplish a sensing task. Various routing Protocols are designed for transmission in WSNs. In this paper, we proposed a hybrid routing protocol: Zonal-Stable Election Protocol (Z-SEP) for heterogeneous WSNs. In this protocol, some nodes transmit data directly to base station while some use clustering technique to send data to base station as in SEP. We implemented Z-SEP and compared it with traditional Low Energy adaptive clustering hierarchy (LEACH) and SEP. Simulation results showed that Z-SEP enhanced the stability period and throughput than existing protocols like LEACH and SEP.

___________________________________________________________________________

## 1. INTRODUCTION

WSNs consist of a large number of sensor nodes that are deployed randomly to monitor physical or environmental conditions, such as temperature, sound, vibration, pressure, motion or pollutants at different locations. Advancement in wireless communications, electronics and technological evolution has enabled the development in the field of WSNs due to their low cost and variety of applications such as health, home and military etc. Research is going on to solve different technical issues in various application areas. Sensor nodes consist of components capable of: sensing data, processing data and also communication components to further transmit or receive data. The protocols and algorithms of such networks must possess self-organizing capabilities to ensure accurate and efficient working of the network.

Communication in WSNs occurs in different ways which totally depends on the application. Generally, there are three main types of communication:
- Clock Driven: Sensors sense and gather data at constantly and periodically communicate.
- Event Driven: Communication is triggered by a particular event.
- Query Driven: Communication occurs in response to a query.

In all three types of communication, efficient use of energy is of concern while studying, designing or deploying such networks to prolong the sensing time and overall lifetime of the network.

Hierarchical routing protocols have been proved more energy efficient routing protocols. Several protocols are designed for homogeneous networks. LEACH [1] is one of the first clustered based routing protocol for homogeneous network. LEACH assigns same probability for all nodes to become cluster head. However, LEACH does not perform well in heterogeneous environment. Heterogeneity of nodes with respect to their energy level has also proved extra lifespan for WSNs. To improve efficiency of WSNs, SEP [2] was proposed. SEP is a two level heterogeneous protocol. SEP assigns different probability (to become cluster head) for nodes on the basis of their energy level. However, SEP does not use extra energy of higher level nodes efficiently.

To send messages from nodes to base station we require minimum dissipation of energy. For such purpose a need of better routing protocol arises which should efficiently utilize energy. Classical approaches were insufficient to fulfill this demand. In this paper we have proposed a hybrid approach for transmitting data to base station. Some nodes send their data directly to base station and some uses clustering algorithm for transmitting data to base station. Our hybrid approach enhanced the stability period, network lifetime and also throughput of the network.

## 2. RELATED WORK AND MOTIVATION

LEACH [1] is a hierarchical clustering algorithm for judicious usage of energy in the network. LEACH uses randomized rotation of the local cluster head. LEACH performs well in homogeneous environment. In LEACH every node has same probability to become a cluster head. However, LEACH is not well suited for heterogeneous environment. SEP is a two level heterogeneous protocol introducing two types of nodes, normal nodes and advance nodes. Advance nodes have more energy than normal nodes. In SEP both nodes (normal and advance nodes) have weighted probability to become cluster head. Advance nodes have more chances to become cluster head than normal nodes. SEP does not guarantee efficient deployment of nodes. Enhanced Stable Election Protocol (E-SEP) [3] was proposed for three level hierarchies. ESEP introduced an intermediate node whose energy lies between normal node and advance node. Nodes elect themselves as cluster head on the basis of their energy level. The drawback of ESEP is same as in SEP. Distributed Energy-Efficient Clustering Protocol (DEEC) [4] shows multilevel heterogeneity. In DEEC the cluster head formation is based on residual energy of node and average energy of the network. In DEEC the high energy node has more chance to become cluster head than low energy node. TEEN [5] is reactive protocol for time critical applications. TEEN was proposed for homogeneous network. In TEEN the criteria for selection of cluster head is same as in LEACH, TEEN introduces hard and soft threshold to minimize the number of transmissions thus saving the energy of nodes. In this way the life span and stability period of the network increases.

In SEP normal nodes and advance nodes are deployed randomly. If majority of normal nodes are deployed far away from base station it consumes more energy while transmitting data which results in the shortening of stability period and decrease in throughput. Hence efficiency of SEP decreases. To remove these flaws we divide network field in regions. As corners are most distant areas in the field, where nodes need more energy to transmit data to base station. So normal nodes are placed near the base station and they transmit their data directly to base station. However advance nodes are deplore far away from base station as they hay more energy. If advance nodes transmit data directly to base station more energy consumes, so to save energy of advance nodes clustering technique is used for advance nodes only.

## 3. TERMINOLOGIES USED

Some basic terminologies we used in the paper are:
- Stability Period: Time interval from the start of the network to the death of the first sensor node.
- Instability Period: Time interval from the death of the first node to the death of the last sensor node.
- Throughput: The total rate of data sent over the network, the rate of data sent from cluster heads to base station as well as the rate of data sent from the nodes to base station.
- Network Lifetime: Time interval from the start of the network to the death of the last alive node.
- Epoch: Number of rounds after which a node becomes eligible for cluster head.
- Data Aggregation: Data collected in sensors are derived from common phenomena so nodes in a close area usually share similar information. A way to reduce energy consumption is data aggregation. Aggregation consists of suppressing redundancy in different data messages. When the suppression is achieved by some signal processing techniques, this operation is called data fusion.

## 4. PROPOSED Z-SEP

In this section we present our proposed protocol. Our protocol is extension of SEP. It follows hybrid approach i.e. direct transmission and transmission via cluster head. Further we discuss in detail the functioning of our protocol.

### 4.1 Network Architecture:

In most routing protocols, nodes are deployed randomly in network field and energy of nodes in network is not utilized efficiently. We modified this theme: network field is divided in three zones: zone 0, Head zone 1 and Head zone 2, on the basis of energy levels and Y co-ordinate of network field.

We assume that a fraction of the total nodes are equipped with more energy. Let *m* be fraction of the total nodes *n*, which are equipped with α time more energy than the other nodes. We refer these nodes as advance nodes, *(1-m)×n* are normal nodes.

Zone 0: Normal nodes are deployed randomly in Zone 0, lying between 20<Y<=80.
Head zone 1: Half of advance nodes are deployed randomly in this zone, lying between 0<Y<=20.
Head zone 2: Half of advance nodes are deployed randomly in Head Zone 2, lying between 80<Y<=100.

The reason behind this type of deployment is that advance nodes have high energy than normal nodes. As corners are most distant places in the field, so if a node is at corner then it requires more energy to communicate with base station so we have deployed high energy nodes (advance nodes) in Head zone 1 and Head zone 2.

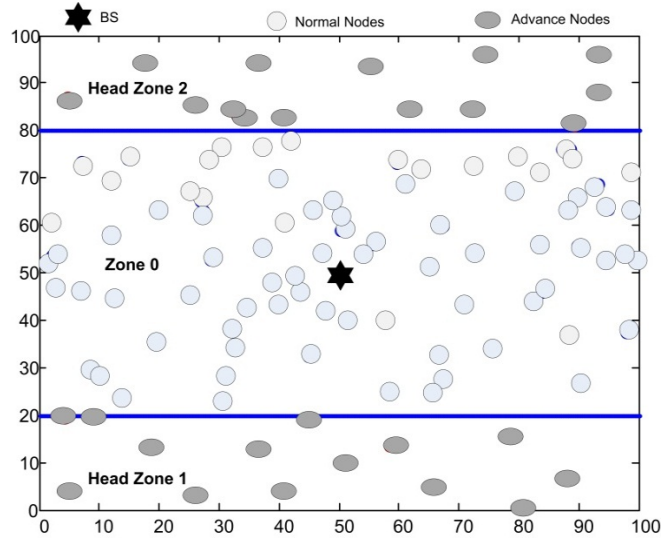

Fig.1 Network Architecture

### 4.2 Z-SEP Operation

Z-SEP uses two techniques to transmit data to base station. Techniques are:
- Direct communication.
- Transmission via Cluster head.

**Direct Communication:**
Nodes in Zone 0 send their data directly to base station. Normal nodes sense environment, gathers data of interest and send it data directly to base station.

**Transmission via Cluster head:**
Nodes in Head zone 1 and Head zone 2 transmit data to base station through clustering algorithm. Cluster head is selected among nodes in Head zone 1 and Head zone 2. Cluster head collect data from member nodes, aggregate it and transmit it to base station. Cluster head selection is most important. As shown in Fig.1 advance nodes are deployed randomly in Head zone 1 and Head zone 2. Cluster is formed only in advance nodes. Assume an optimal number of clusters *Kopt* and *n* is the number of advance nodes. According to SEP optimal probability of cluster head is

$$Popt = \frac{Kopt}{n} \qquad (1)$$

Every node decides whether to become cluster head in current round or not. A random number between 0 and 1 is generated for node. If this random number is less than or equal threshold *T(n)* for node then it is selected as cluster head. Threshold *T(n)* is given by

$$T(n) = \begin{cases} \dfrac{Popt}{1 - Popt\left(r \times mod\, \dfrac{1}{Popt}\right)} & if\ n \in G \\ 0 & otherwise \end{cases} \qquad (2)$$

Where G is the set of nodes which have not been cluster heads in the last 1/Popt rounds.
Probability for advance nodes to become cluster head is proposed in [2] which is

$$Padv = \frac{Popt}{1+(\alpha.m)} \times (1+\alpha) \qquad (3)$$

Accordingly the threshold for advance nodes is

$$T(adv) = \begin{cases} \dfrac{Padv}{1 - Padv\left(r \times mod\dfrac{1}{Padv}\right)} & if\ adv \in G' \\ 0 & otherwise \end{cases} \qquad (4)$$

G' is the set of advance nodes that have not been cluster head in the last 1/Padv rounds.

Once the cluster head is selected then the cluster head broadcasts an advertisement message to the nodes. The nodes receive the message and decide to which cluster head it will belong for the current round. This phase is called as *cluster formation phase.*

On the basis of received signal strength, nodes respond to cluster head and become member of cluster head. Cluster head then assign a TDMA schedule for the nodes during which nodes can send data to cluster head. After the clusters formation, every node data and sends it to the cluster head in the time slot allocated by the cluster head to the node. This phase is shown in Fig. 2.

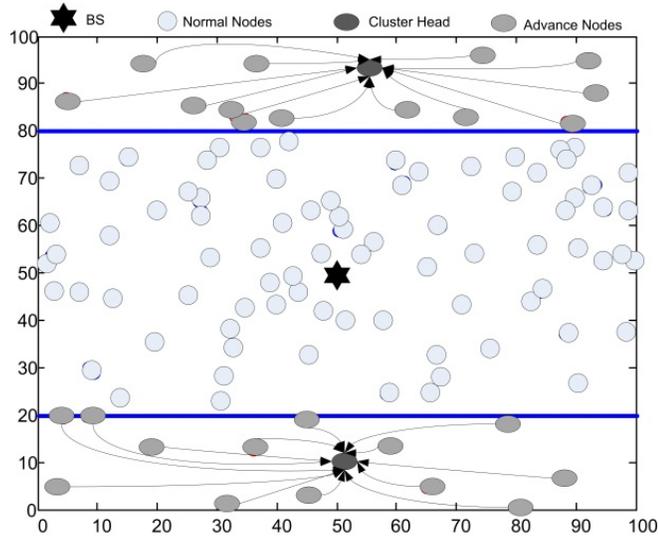

Fig.2 Nodes sending data to cluster head

When data is received from nodes, Cluster head then aggregates this data and send it to the base station this phase is called as *transmission phase*. Fig.3 illustrates this phase.

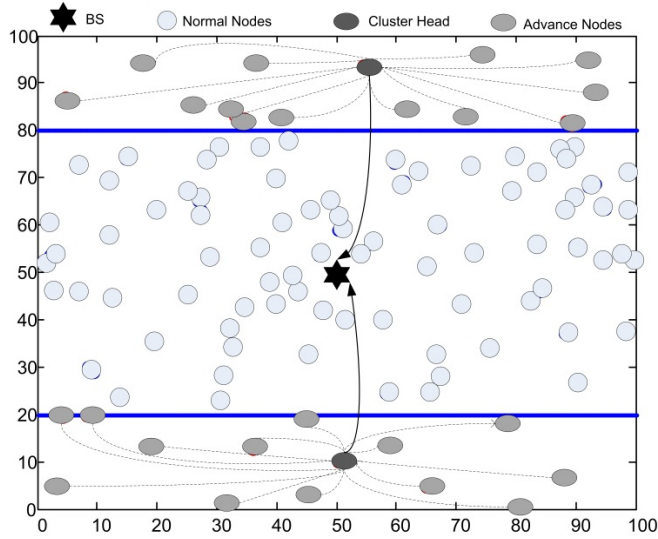

Fig.3 Cluster head transmitting data to base station

The reason why normal nodes (deployed in Zone 0) do not form cluster is because energy of normal node is less than advance node, and cluster head consumes more energy than cluster members in receiving data from cluster members. If we allow normal nodes to become cluster head they die soon resulting in the shortening of stability period. Fig.4 illustrates Z-SEP operation.

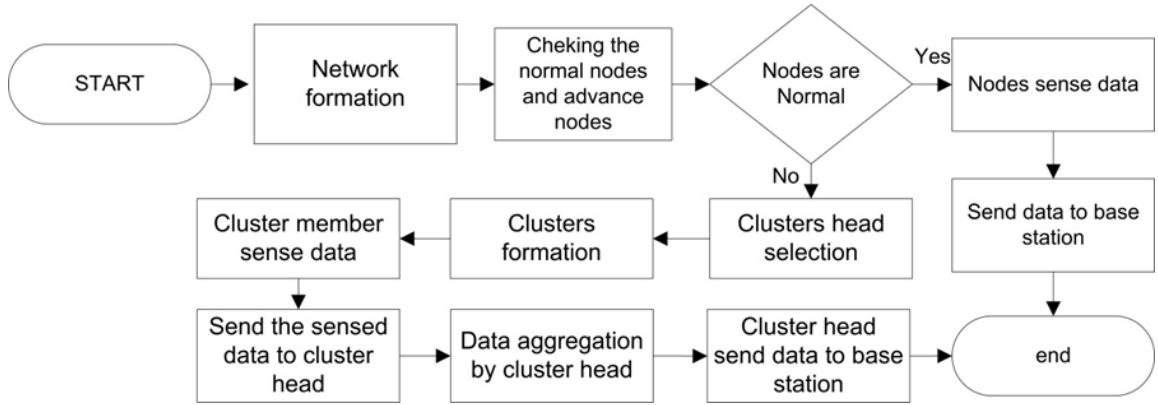

Fig.4 Flow chart of Z-SEP

## 5. SIMULATIONS

We simulate our proposed protocol in a field with dimensions 100m×100m and 100 nodes deployed in specific zones with respect to their energy. Base station is placed in the center of the network field. We are using the first order radio model as used in SEP. MATLAB is used to implement the simulations.
Specifically, we have following settings.

Let 20% of nodes are advance nodes and half of them are deployed in Head zone 1 and half in Head zone 2. since *Popt* is 0.1 so we have 2 cluster heads per round. One cluster head in Head zone 1 and one in Head zone 2 per round.

Other simulation parameters are shown in Table 1.

Table 1: Simulation parameters

| Parameters | Value |
| --- | --- |
| Initial energy $E_o$ | 0.5 J |
| Initial energy of advance nodes | $E_o(1+\alpha)$ |

| Energy for data aggregation $E_{DA}$ | 5 nJ/bit/signal |
|---|---|
| Transmitting and receiving energy $E_{elec}$ | 5 nJ/bit |
| Amplification energy for short distance $E_{fs}$ | 10 Pj/bit/m$^2$ |
| Amplification energy for long distance $E_{amp}$ | 0.013 pJ/bit/m$^4$ |
| Probability $P_{opt}$ | 0.1 |

## 6. RESULT AND DISCUSSION

Here, we compare the results of our protocol with SEP and LEACH. We have introduced heterogeneity in LEACH, with the same setting as in our proposed protocol, so as to access the performance of all the protocol in presence of heterogeneity. Our goals in conducting simulation are
- To examine the stability period of LEACH, SEP and Z-SEP.
- We also examine the throughput of LEACH, SEP and Z-SEP.

Fig.5 and Fig.6 shows result for the case when $m=0.2$ and $\alpha=1$. This means that there are 20 advance nodes out of total nodes which are 100. According to our proposed protocol 10 advance nodes will be deployed randomly in Head zone 1 and 10 advance nodes will be placed in Head zone 2.

Fig.5 shows the number of alive nodes against rounds. Fig.5 clearly shows that our protocol is enhanced from SEP and LEACH in terms of stability. As LEACH is very sensitive to heterogeneity so nodes die at a faster rate. SEP performs better than LEACH in two level heterogeneity, because SEP has weighted probability for selection of cluster head for both normal nodes and advance nodes. Z-SEP performs better than LEACH and SEP, because nodes in Zone 0 (normal nodes) communicates directly to base station while nodes in head zone 1 and head zone 2 communicates via cluster head to base station: As in clustering technique, cluster head consumes energy in the form of data aggregation and also by receiving data from nodes in the cluster. So this energy is conserved in normal nodes as they do not have to aggregate data and receive data from other nodes, so energy is not dissipated as that of cluster head, resulting the increase of stability period. In Fig.5, we can see that network lifetime is also increased because of the advance node. Advance nodes have α time more energy than normal nodes so advance nodes die later than normal nodes. So this increases the instability period.

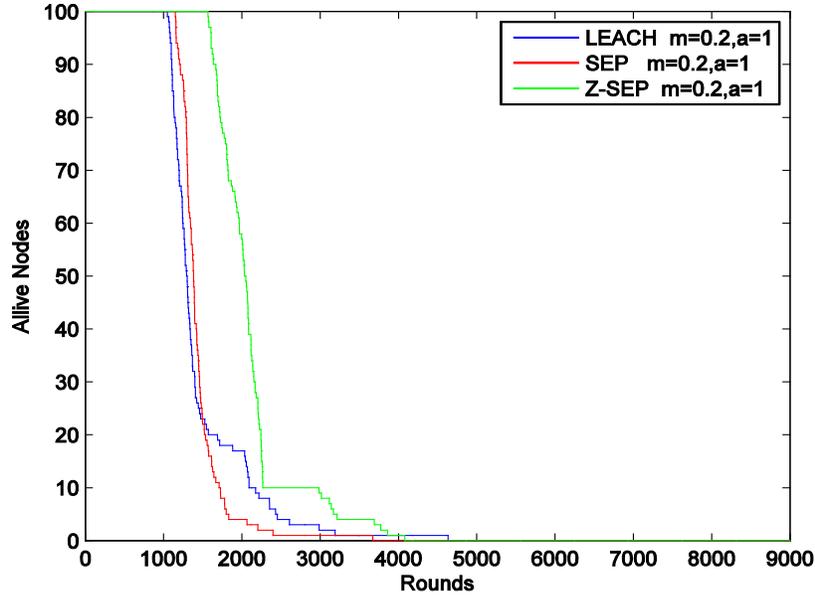

Fig.5 Alive nodes in LEACH, SEP and Z-SEP

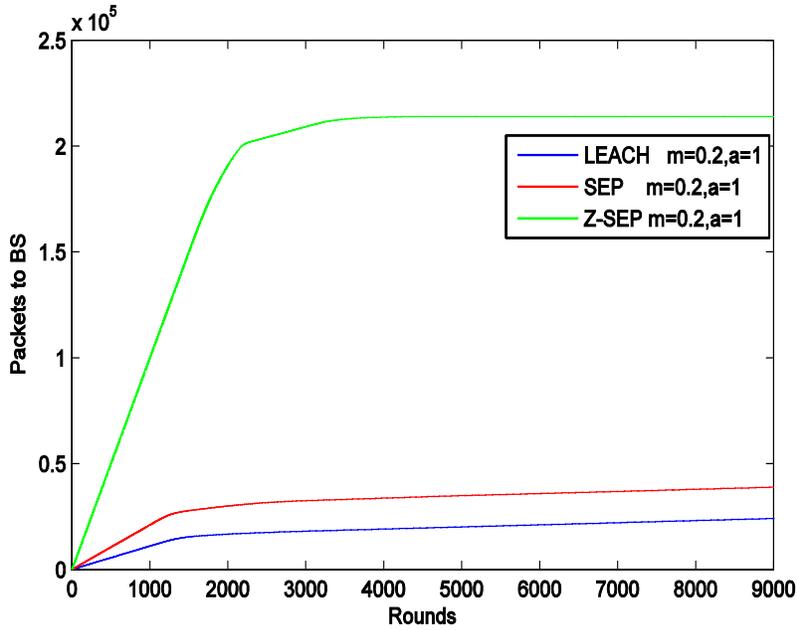

Fig.6 Throughput of LEACH, SEP and Z-SEP

In Fig.6, we can see that throughput of Z-SEP is far better than LEACH and SEP because every normal directly send data to base station. Throughput of LEACH and SEP is less than Z-SEP because only cluster head send data to base station.

Fig.7 and Fig.8 shows result for the case when $m$=0.1 and $α$=2. We have total 10 advance nodes in the field, 5 nodes in Head zone 1 and 5 nodes in Head zone 2. However there energy is increased i.e. $α$=2.

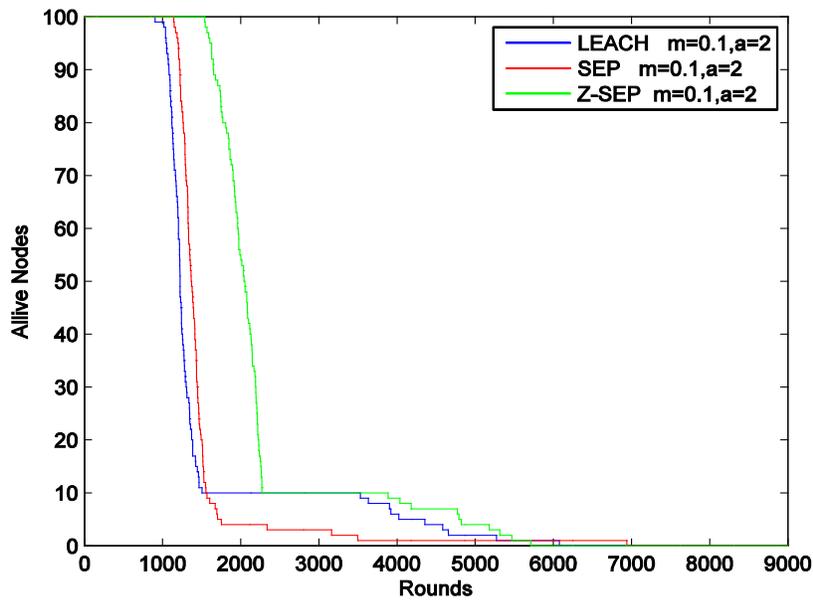

Fig.7 Alive nodes in LEACH, SEP and Z-SEP

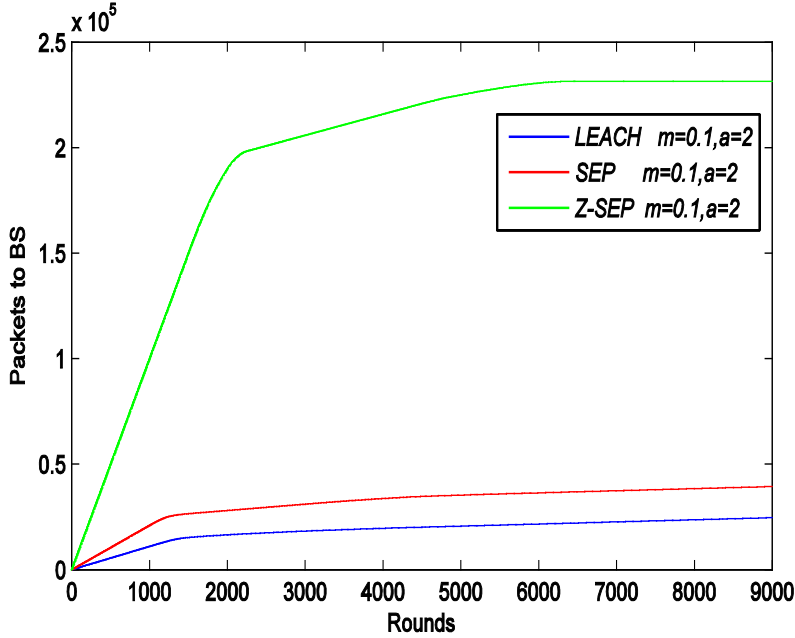

Fig.8 Throughput of LEACH, SEP and Z-SEP

From Fig.7, we can see that stability period of Z-SEP is almost same for both cases i.e. ($m$=0.2, $\alpha$=1 and $m$=0.1, $\alpha$=2). The reason behind is that normal nodes have same amount of energy, they consume same amount of energy and they die almost at the same time as before, however network lifetime is increased because of the extra energy of advance nodes. Stability period of LEACH is decreased because LEACH is very sensitive to heterogeneity. LEACH does not have weighted probability as in SEP for even distribution of extra energy. In LEACH every node has equal chance to become cluster head so normal nodes die sooner than advance nodes.

Fig. 8 shows the throughput of LEACH, SEP and Z-SEP. Throughput of Z-SEP is greater than LEACH and SEP although energy of advance node has been increased.

Table 2: Comparison Table When $m$=0.2 and $\alpha$=1

| Protocol | Stability Period (Rounds) | Network Lifetime (Rounds) | Throughput (Packets) |
|---|---|---|---|
| LEACH | 1018 | 4685 | $1.99\times10^4$ |
| SEP | 1089 | 3005 | $3.43\times10^4$ |
| Z-SEP | 1531 | 4119 | $2.21\times10^5$ |

Table 3: Comparison Table When $m$=0.1 and $\alpha$=2

| Protocol | Stability Period (Rounds) | Network Lifetime (Rounds) | Throughput (Packets) |
|---|---|---|---|
| LEACH | 899 | 5583 | $2.44\times10^4$ |
| SEP | 1150 | 5078 | $4.02\times10^4$ |
| Z-SEP | 1584 | 5966 | $2.26\times10^5$ |

In Table 2 and Table 3, we have compared the average results for LEACH, SEP and Z-SEP. Approximately 50% stability period of our proposed protocol is increased from LEACH and SEP, however network lifetime is decreased when compared with LEACH. When compared with SEP, Z-SEP network life time is increased due to advance nodes which die slower than normal nodes. Network lifetime of SEP is short because of the weighted probability for normal and advance nodes in the field.

## 7. CONCLUSION

In this paper, we proposed Z-SEP for heterogeneous environment: two level heterogeneity. The field is divided in to three zones: Zone 0, Head Zone 1 and Head Zone 2. Normal nodes are only deployed in zone 0 to reduce the energy consumption and they transmit data directly to base station. Half of advanced nodes are deployed in Head zone 1 and half in Head zone 2 and they use clustering technique to transmit data to base station. Results have shown that the stability period is increased approximately 50%, by just altering the deployment of the different type of nodes in different zones according to their energy requirement. Throughput of Z-SEP is also increased compared with LEACH and SEP.